\documentclass[a4paper,14pt]{article}

\usepackage{multicol}
\usepackage{amsthm}

\usepackage[dvipdfmx]{graphicx}
\usepackage{amssymb}
\usepackage{amsmath}
\usepackage{color}

\begin{document}

\fontsize{14pt}{16.5pt}\selectfont

\begin{center}
\bf{Ultradiscrete Bifurcations for One Dimensional Dynamical Systems}
\end{center}
\fontsize{12pt}{11pt}\selectfont
\begin{center}
Shousuke Ohmori$^{*)}$ and Yoshihiro Yamazaki\\ 
\end{center}

\noindent
\it{Department of Physics, Waseda University, Shinjuku, Tokyo 169-8555, Japan}\\

\noindent
*corresponding author: 42261timemachine@ruri.waseda.jp\\
~~\\
\rm
\fontsize{11pt}{14pt}\selectfont\noindent

\baselineskip 30pt

\noindent
{\bf Abstract}\\
Bifurcations of one dimensional dynamical systems are discussed 
based on some ultradiscrete equations.
The ultradiscrete equations are derived 
from normal forms of one-dimensional nonlinear differential equations, 
each of which has saddle-node, transcritical, 
or supercritical pitchfork bifurcations.
An additional bifurcation, which is similar to flip bifurcation, 
is found in ultradiscrete equations 
for supercritical pitchfork bifurcation.
Dynamical properties of these ultradiscrete bifurcations can be characterized 
with graphical analysis. 
As an example of application of our treatment, 
we focus on an ultradiscrete equation of FitzHugh-Nagumo model, 
and discuss its dynamical properties.\\


\section{Introduction}
\label{sec:1}

%
Ultradiscretization is a limiting procedure 
converting a difference equation into other type of difference equation 
subject to max-plus algebra\cite{Tokihiro1996}.
In this procedure, first, a positive variable $u_n$ of a difference equation 
is transformed into $U_{n}$ 
by $u_n=\exp(U_n/\varepsilon )$, where $\varepsilon $ is a positive parameter. 
After the transformation, the formulae 
	\begin{eqnarray}
\begin{cases}
		\displaystyle\lim_{\varepsilon  \to +0} \varepsilon  \log(e^{A/\varepsilon }+e^{B/\varepsilon }+\cdot \cdot \cdot )~=~\max(A,B,\dots),\\
		\displaystyle\lim_{\varepsilon  \to +0} \varepsilon  \log(e^{A/\varepsilon }\cdot e^{B/\varepsilon }\cdot \dots ) = A+B+ \dots,
\end{cases}
		\label{eqn:1-1}
	\end{eqnarray}
are adopted. 
Then, we obtain a new difference equation 
called an ultradiscrete equation of $U_n$. 
This limiting procedure brings piecewise linearization 
of the original equation, 
and an obtained ultradiscrete equation can be associated 
with a cellular automaton.
Ultradiscretization has been successfully applied 
to integrable systems\cite{Grammaticos1997}.
Actually, some cellular automata have been derived 
from soliton equations such as KdV equation 
through ultradiscretization\cite{Tokihiro1996,Takahashi1990,Matsukibara1997,Nagai1999}.
Lately, ultradiscretization has been applied to non-integrable non-equilibrium dissipative systems such as reaction diffusion systems, which are also expressed by non-linear differential equations\cite{Nagatani1998,Murata2013,Ohmori2014,Matsuya2015,Murata2015,Gibo2015,Ohmori2016}. 

For discretization of non-linear differential equations, 
Murata proposed the tropical discretization\cite{Murata2013}.
Now we consider the equation of $u$ : 
	\begin{eqnarray}
		\displaystyle~\frac{d u}{d t} = f(u)-g(u).
		\label{eqn:1-2}
	\end{eqnarray}
Here, $u=u(t)>0$ and $f(u),g(u)\geq 0$.
The tropical discretization leads eq.(\ref{eqn:1-2}) 
to the discrete equation 
	\begin{eqnarray}
		u _{n+1}=u _n \frac{u _n+\Delta tf(u _n)}{u _n+\Delta tg(u _n)},
		\label{eqn:1-3}
	\end{eqnarray}
where $\Delta t$ is the discretized time interval.
$n$ shows the number of iteration steps and is non-negative integer;
$u_n = u(n \Delta t)$.
In order to obtain an ultradiscrete equation 
for eq.(\ref{eqn:1-2}), 
the variable transformations are adopted 
to eq.(\ref{eqn:1-3}): 
$\Delta t = e^{T/\varepsilon }$, $u_n = e^{U_n/\varepsilon }$, 
$f(u_n)=e^{F(U_n)/\varepsilon} $, 
and $~g(u_n)=e^{G(U_n)/\varepsilon }$.
After these transformation, 
we obtain the following ultradiscrete equation 
by eq.(\ref{eqn:1-1}): 
	\begin{eqnarray}
		U_{n+1}=U_n+\max \{U_n,T+F(U_n)\}-\max \{U_n, T+G(U_n)\}.
		\label{eqn:1-4}
	\end{eqnarray}
%
%

So far, relationship between solutions 
of original non-linear differential equations 
and those of their ultradiscrete equations 
has also been discussed for non-integrable dissipative systems
\cite{Murata2013,Ohmori2014,Matsuya2015,Ohmori2016}.
Meanwhile, these differential equations have been widely studied 
from the view point of dynamical systems\cite{Prigogine,Guckenheimer,Strogatz,Nicolis}.
For instance, stability of fixed points and bifurcation phenomena 
have been treated. 
And frameworks for their treatments have been established.  
We feel that application of ultradiscretization to the nonlinear dynamical systems 
is important but seems to be insufficient now.

In this paper, we discuss dynamical properties 
of some ultradiscrete equations 
for the non-linear dynamical systems.
Especially, we focus on normal forms of saddle node, transcritical,  
and supercritical pitchfork bifurcations \cite{Guckenheimer,Strogatz}.
In the next section, ultradiscrete equations of these normal forms are derived.
%
%
It is noted that ultradiscrete bifurcations, 
namely bifurcations of ultradiscrete equations, are characterized 
by the piecewise linearity with max-plus algebra\cite{Baccelli}. 
To elucidate essence of the ultradiscrete bifurcations, 
we introduce some simple max-plus discrete equations in Section 3. 
These equations have the same piecewise linearity as the ultradiscrete equations.
By introducing the simple max-plus discrete equations, 
it is easy to sketch the piecewise linear graphs 
and easy to grasp dynamical properties of the bifurcations; 
this approach is known as a graphical analysis\cite{Devaney}.
Discussion and conclusion are given in Sections 4 and 5, respectively.
%
%

\section{Ultradiscretization of one dimensional normal forms}
\label{sec:2}

The one-dimensional normal forms of the saddle-node, transcritical, 
and supercritical pitchfork bifurcations are generally  given 
as the following nonlinear equations, respectively \cite{Guckenheimer,Strogatz};
\begin{eqnarray}
  \text{(saddle-node)} & &
  \frac{du}{dt} = r+u^2,
  \label{eqn:2-1a}\\
  \text{(transcritical)} & &
  \frac{du}{dt} = ru-u^2,
  \label{eqn:2-1b}\\
  \text{(supercritical pitchfork)} & &
  \frac{du}{dt} = ru-u^3,
  \label{eqn:2-1c}
\end{eqnarray}
where $r$ is a bifurcation parameter.
The bifurcation points of these three normal forms are $r = 0$.
In this section, we first derive ultradiscrete equations 
from equations for these bifurcations.
Then, we show the bifurcation properties 
of the obtained ultradiscrete equations.

\subsection{Saddle-node bifurcation}
\label{sec:2-1}

Now we focus on eq.(\ref{eqn:2-1a}), the normal form of saddle-node bifurcation.
In eq.(\ref{eqn:2-1a}),  there is no fixed point for $r > 0$. 
For $r<0$, the two fixed points exist; 
one of them is positive and another one is negative.
Note that, in general, existence of negative values 
makes it difficult to apply ultradiscretization. 
To avoid this difficulty for eq.(\ref{eqn:2-1a}), we consider the following equation 
instead of eq.(\ref{eqn:2-1a}).
%
%
%
%
%
%
	\begin{eqnarray}
	\frac{du}{dt}  =  c+u(u-2) .
	\label{eqn:2-1-1a}
	\end{eqnarray}
%
%
$c$ is the bifurcation parameter, and 
the saddle-node bifurcation occurs at $c= 1$.
If $c<1$,  eq.(\ref{eqn:2-1-1a}) has the two fixed points 
$u_{-}$ and $u_{+}$, 
where $u_{\pm}={1 \pm \sqrt{ 1-c}}$. 
Note that $u = u_{-}$ and $u_{+}$ are stable and unstable, respectively.
If $c=1$, $u=1$ is half-stable.
If $c>1$, there is no fixed point.

By using the tropical discretization, 
the discrete equation of eq.(\ref{eqn:2-1-1a}) is obtained as 
	\begin{eqnarray}
		u_{n+1}=\displaystyle\frac{u_n+{\Delta t}[(u_n)^2+c]}{1+2{\Delta t}}.
	\label{eqn:2-1-2}
	\end{eqnarray}
It is confirmed that eq.(\ref{eqn:2-1-2}) shows the saddle-node bifurcation 
by changing the parameter $c$ 
and that the bifurcation point $c=1$ of eq.(\ref{eqn:2-1-2}) is the same 
as that of eq.(\ref{eqn:2-1-1a}).
After the variable transformations 
	\begin{equation}
		\Delta t = e^{T/\varepsilon},~~~ u_n = e^{U_n/\varepsilon},~~~ c =e^{C/\varepsilon },
		\label{eqn:2-1-3}
	\end{equation}
and ultradiscretization by eq.(\ref{eqn:1-1}), 
we obtain the ultradiscrete equation
	\begin{eqnarray}
		U_{n+1} = \max\{U_n, T+ \max(2U_n, C)\}-\max \{0, T\}.
	\label{eqn:2-1-4}
	\end{eqnarray}
Assuming $T\geq \max\{0,-C/2\}$, 
the following ultradiscrete equation can be derived 
from eq.(\ref{eqn:2-1-4}):
	\begin{eqnarray}
		U_{n+1}=\max (2U_n, C).
	\label{eqn:2-1-5}
	\end{eqnarray}
(For derivation of eq.(\ref{eqn:2-1-5}), see Appendix \ref{sec:6-1}.)
Dynamical properties of $U_n$ in eq.(\ref{eqn:2-1-5}) are summarized as {\bf 2.1(i)} $\sim$ {\bf 2.1(iii)}
depending on the value of $C$ (see Appendix \ref{sec:7-1} for details).

\begin{description}
	\item[2.1(i)] When $C>0$, $U_{n+1}>U_n$ for any $n$; 
		there is no fixed point.
	\item[2.1(ii)] When $C=0$, $U_n=0$ is a fixed point.
	If $U_n < 0$, $U_{n+1}=0$. If $U_n> 0$, $U_{n+1}>U_n$.
	Hence, $U_n=0$ is half-stable.
	\item[2.1(iii)] We set $C<0$.
	$U_n=C$ and $U_n=0$ are fixed points. \\
         (a) If $U_n <  C$ or $C < U_n \leq C/2$, $U_{n+1}=C$. \\
         (b) If $C/2<U_n<0$, $C< U_{n+1}<U_n$. \\
         (c) If $U_n>0$, $U_{n+1}>U_n$. \\
         Hence, $U_n=C$ is stable and $U_n=0$ is unstable.
\end{description}
From these dynamical properties of $U_n$, 
the ultradiscrete bifurcation point $C=0$ corresponds 
to the bifurcation point $c=1$ for eq.(\ref{eqn:2-1-1a}), 
and {\bf 2.1(i)}, {\bf (ii)}, and {\bf (iii)} correspond 
to the cases of $c>1$, $c=1$, and $c<1$ in eq.(\ref{eqn:2-1-1a}), 
respectively. 
%
%
Thus, the ultradiscrete equation (\ref{eqn:2-1-5}) reproduces a similar saddle-node bifurcation 
to the original normal form.
%
%
%
%
%


\subsection{Transcritical bifurcation}
\label{sec:2-2}

Next we focus on eq. (\ref{eqn:2-1b}), the normal form for the transcritical bifurcation.
To apply ultradiscretization, we consider
%
%
%
%
%
%
%
	\begin{eqnarray}
	\frac{du}{dt} = (u-1)(c -u).
	\label{eqn:2-2-1a}
	\end{eqnarray}
%
%
In eq.(\ref{eqn:2-2-1a}), $c$ is the bifurcation parameter 
and the transcritical bifurcation occurs at $c = 1$.
By the tropical discretization for eq.(\ref{eqn:2-2-1a}), we obtain
	\begin{eqnarray}
		u_{n+1}=u_n\displaystyle
		\frac{u_n+{\Delta t}(1 + c) u_n}{u_n+{\Delta t}[(u_n)^2+ c]}.
	\label{eqn:2-2-2}
	\end{eqnarray}
Then ultradiscretization of eq.(\ref{eqn:2-2-2}) produces
	\begin{eqnarray}
		U_{n+1}=U_n+\max\{U_n,T+U_n+\max(0,C)\}-\max\{U_n,T+\max(2U_n,C)\},
	\label{eqn:2-2-3}
	\end{eqnarray}
where we set the variable transformations 
	\begin{equation}
			\Delta t = e^{T/\varepsilon},~~~ u_n = e^{U_n/\varepsilon},~~~  c =e^{C/\varepsilon }.
	\label{eqn:2-2-4}
	\end{equation}
Now we assume $T \geq -C/2$.
Then, eq.(\ref{eqn:2-2-3}) can be simplified as follows.
	\begin{eqnarray}
		U_{n+1}=2U_n + \max(0, C)-\max(2U_n,C).
	\label{eqn:2-2-5}
	\end{eqnarray}
(For derivation of eq.(\ref{eqn:2-2-5}), see Appendix \ref{sec:6-2}.)
Dynamical properties of $U_n$ given by eq.(\ref{eqn:2-2-5}) are summarized 
as {\bf 2.2(i)} $\sim$ {\bf (iii)} (see Appendix \ref{sec:7-2} for further explanation).

\begin{description}
	\item[2.2(i)] When $C>0$, $U_n=0$ and $U_n=C$ are fixed points. \\
    (a) If $U_n<0$, $U_{n+1}<U_n$. \\
    (b) If $0<U_n<\frac{C}{2}$, $U_{n+1}>U_n$. \\
    (c) If $U_n>C$ or $\frac{C}{2}\leq U_n <C$, $U_{n+1}=C$. \\
	Hence, $U_n=0$ is unstable and $U_n=C$ is stable.
	\item[2.2(ii)] When $C=0$, $U_n=0$ is a fixed point. \\
    (a) If $U_n>0$, $U_{n+1}=0$. \\
    (b) If $U_n<0$, $U_{n+1}<U_n$. \\
    Then, $U_n=0$ is the half-stable point.
	\item[2.2(iii)] When $C<0$, $U_n=0$ and $U_n=C$ are fixed points. 
    As in the case of {\bf 2.2(i)}, \\
    (a) If $U_n<C$, $U_{n+1}<U_n$. \\
    (b) If $C<U_n<\frac{C}{2}$, $U_{n+1}>U_n$. \\
    (c) If $U_n>0$ or $\frac{C}{2} \leq U_n < 0$, $U_{n+1}=0$. \\
   Hence, $U_n=0$ is stable and $U_n=C$ is unstable. 
\end{description}
These dynamical properties are similar 
to those of the transcritical bifurcation by eq.(\ref{eqn:2-2-1a}).
Actually, the ultradiscrete bifurcation point $C=0$ corresponds 
to the bifurcation point $c=1$ for eq.(\ref{eqn:2-2-1a}), 
and {\bf 2.2(i)}, {\bf (ii)}, and {\bf (iii)} correspond 
to the cases of $c>1$, $c=1$, and $c<1$ in eq.(\ref{eqn:2-2-1a}), 
respectively. 
%
%


\subsection{Supercritical pitchfork bifurcation}
\label{sec:2-3}


For ultradiscretization, 
let us consider the following equation instead of eq.(\ref{eqn:2-1c}), 
\begin{equation}
	\frac{dv}{dt} = r(v + v^{2}) -v^3.
	\label{eqn:2-3-0}
\end{equation}
After the variable transformations $v = u - 1$ and $r = 3(c-1)$ 
to eq.(\ref{eqn:2-3-0}), we obtain
	\begin{eqnarray}
	\frac{du}{dt} = 3cu(u-1)-u^3+1, 
	\label{eqn:2-3-1}
	\end{eqnarray}
where $c$ is positive.
$c$ is the bifurcation parameter and supercritical pitchfork bifurcation occurs at $c=1$.
The discrete equation of eq.(\ref{eqn:2-3-1}) 
by the tropical discretization is
	\begin{eqnarray}
		u_{n+1}=\displaystyle\frac{u_n+{\Delta t}[3c(u_n)^2 + 1]}{1+{\Delta t}[(u_n)^2+ 3c]}.
	\label{eqn:2-3-2}
	\end{eqnarray}
Setting the variable transformations
	\begin{equation}
			\Delta t = e^{T/\varepsilon},~~~ u_n = e^{U_n/\varepsilon},~~~  c =e^{C/\varepsilon },
	\label{eqn:2-3-3}
	\end{equation}
we obtain the ultradiscrete equation
	\begin{eqnarray}
		U_{n+1}=\max\{U_n,T+\max(2U_n+C,0)\}-\max\{0,T+\max(2U_n,C)\}.
	\label{eqn:2-3-4}
	\end{eqnarray}
Assuming $T \geq \max(-C,0)$, 
the following ultradiscrete equation is obtained from eq.(\ref{eqn:2-3-4}).
	\begin{eqnarray}
		U_{n+1}=\max(2U_n+C, 0)-\max(2U_n,C).
	\label{eqn:2-3-5}
	\end{eqnarray}
(For derivation of eq.(\ref{eqn:2-3-5}), see Appendix \ref{sec:6-3}.)
$C$-dependence of dynamical properties of eq.(\ref{eqn:2-3-5}) 
are summarized as follows (see Appendix \ref{sec:7-3} for details);

\begin{description}
	\item[2.3(i)] When $C >0$, the following three values of $U_{n}$ are fixed points:
    $U_n = $ $-C$, $0$, and $C$. \\
    (a) If $U_n<-C$ or $-C< U_n \leq -\frac{C}{2}$, $U_{n+1}=-C$. \\
    (b) If $-\frac{C}{2} < U_n <0$, $U_{n+1}<U_n$. \\
    (c) If $0<U_n<\frac{C}{2}$, $U_{n+1}>U_n$. \\
    (d) If $U_n>C$ or $\frac{C}{2} \leq U_n <C$, $U_{n+1}=C$. \\
	From (a) $\sim$ (d), $U_{n} = -C$ and $C$ are stable, and $U_{n} = 0$ is unstable. 
	\item[2.3(ii)] When $C = 0$, $U_{n} = 0$ is the only one fixed point. 
    If $U_n \not = 0$, $U_{n+1}=0$.
    Hence, $U_{n} = 0$ is stable.
	\item[2.3(iii)] When $C <0$, $U_{n} = 0$ is the only one fixed point. \\
    (a) If $U_n \leq \frac{C}{2}$, $U_{n+1}=-C$. \\
    (b) If $\frac{C}{2} < U_n <0$, $U_{n+1}>-U_n$. \\
    (c) If $0<U_n<-\frac{C}{2}$, $U_{n+1}<-U_n$. \\
    (d) If $U_n\geq -\frac{C}{2}$, $U_{n+1}=C$. \\
	Therefore, $U_{n} = 0$ is unstable. 
\end{description}
Focusing on {\bf 2.3(i)} and {\bf 2.3(ii)}, 
it seems that eq.(\ref{eqn:2-3-5}) exhibits a bifurcation 
similar to the supercritical pitchfork bifurcation.
However, eq.(\ref{eqn:2-3-1}) does not 
possess a dynamical property like {\bf 2.3(iii)}; 
the dynamical transition between {\bf 2.3(ii)} and {\bf 2.3(iii)}
is rather similar to the flip bifurcation\cite{Kuznetsov}.
%

%


\section{Graphical analysis}
\label{sec:3}

In the previous section, we have shown the bifurcation properties 
of the ultradiscrete equations (\ref{eqn:2-1-5}),(\ref{eqn:2-2-5}), and (\ref{eqn:2-3-5}) 
with the bifurcation parameter $C$.
In this section, we propose some general max-plus discrete equations, 
which exhibit the same properties of the ultradiscrete equations. 
Especially, by using a graphical analysis, their dynamical properties can be visualized.
The graphical analysis is well known as a method to intuitively understand one dimensional discrete iterated dynamics\cite{Devaney, Robinson, Kuznetsov}.
In general, time evolution of $U_{n}$ can be described
as $U_{n+1}=f(U_n, U_{n-1}, \cdots; C)$.
Here we consider the case where $U_{n+1}$ is determined only by $U_{n}$: 
$U_{n+1}=f(U_n; C)$.
%
%
%
%
%

\subsection{Saddle-node bifurcation}
\label{sec:3-1}

\begin{figure}[h!]
	\begin{center}
	\includegraphics[bb=0 0 960 720, width=5cm]{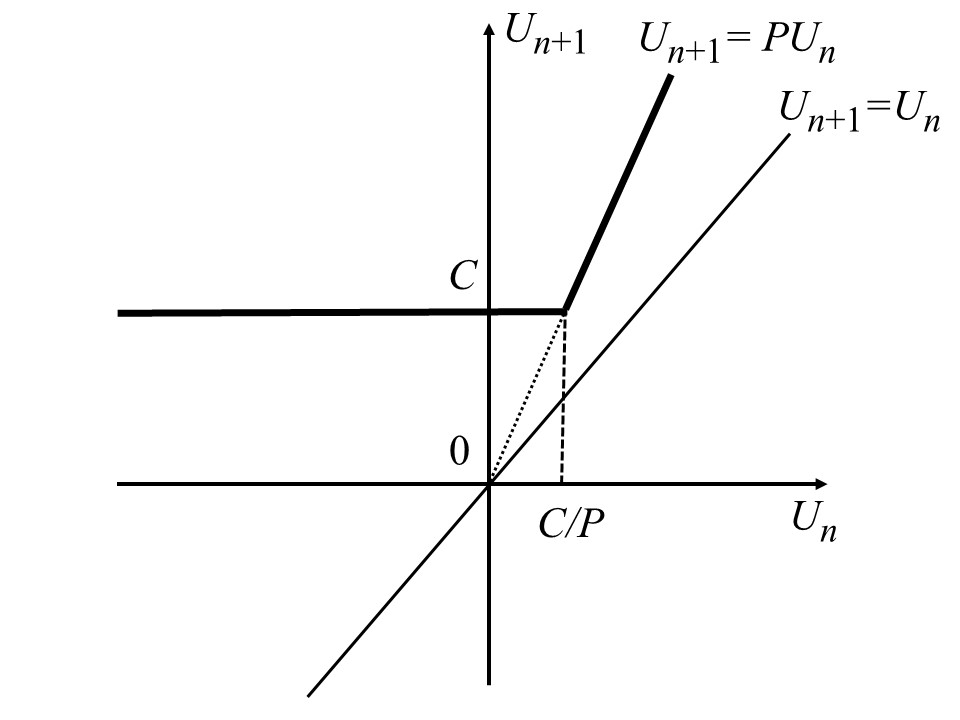}
	\hspace{0mm}
	\includegraphics[bb=0 0 960 720, width=5cm]{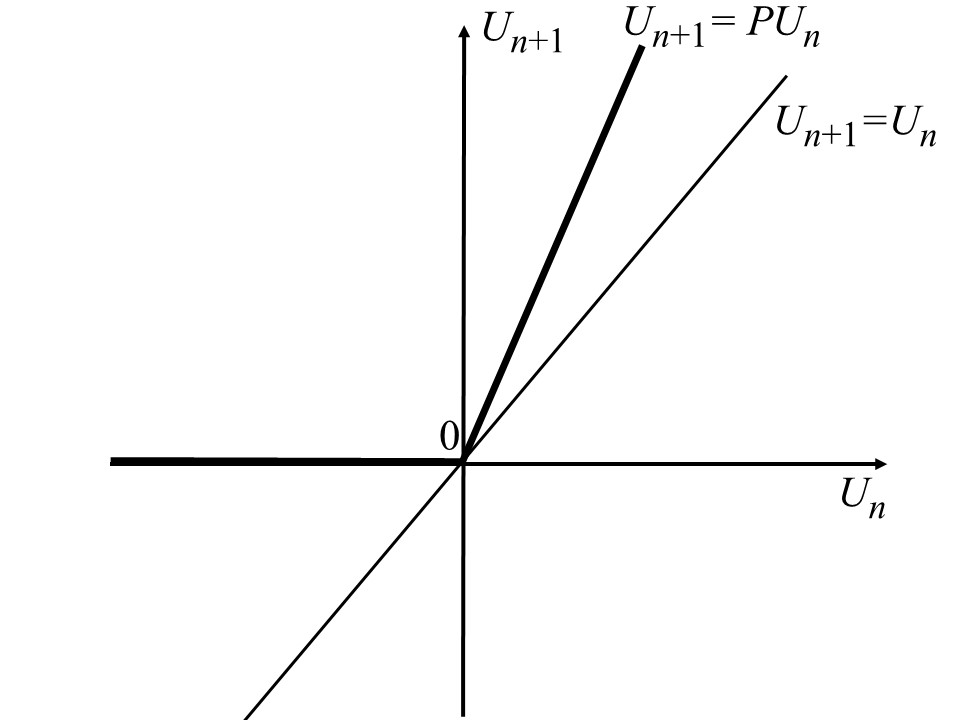}
	\hspace{0mm}
	\includegraphics[bb=0 0 960 720, width=5cm]{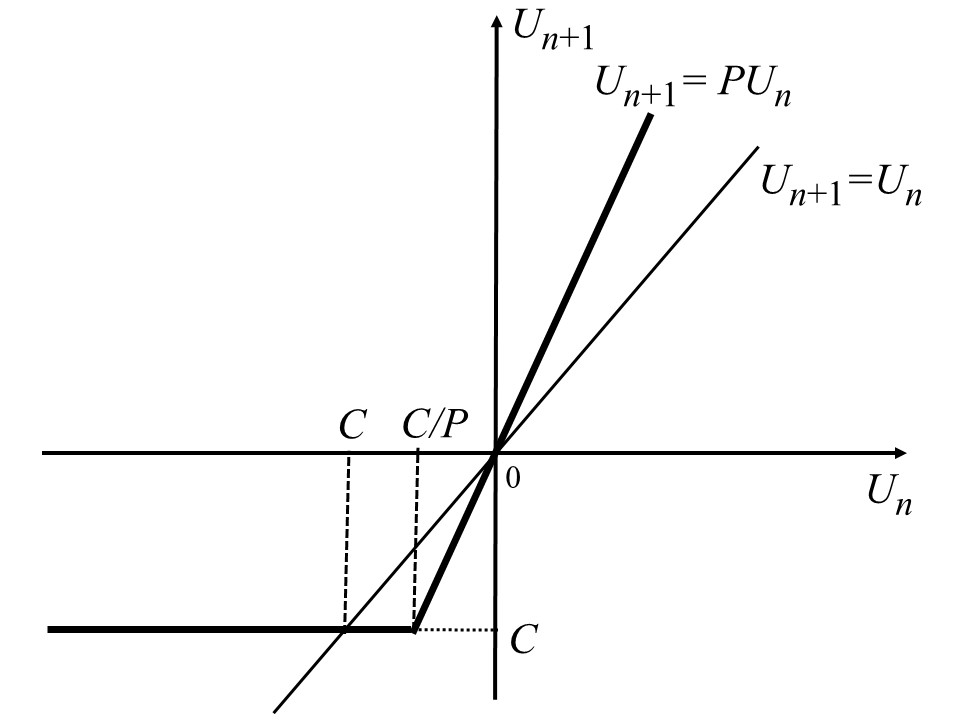}
	\\
	(a)
	\hspace{4cm}
	(b)
	\hspace{4cm}
	(c)
	\caption{\label{Fig.1} The graphs of eq.(\ref{eqn:3-1}). (a) $C>0$, (b) $C=0$, and (c) $C<0$.}
	\end{center}
\end{figure}

\begin{figure}[h!]
\begin{center}
\includegraphics[bb=0 0 960 720, width=5.5cm]{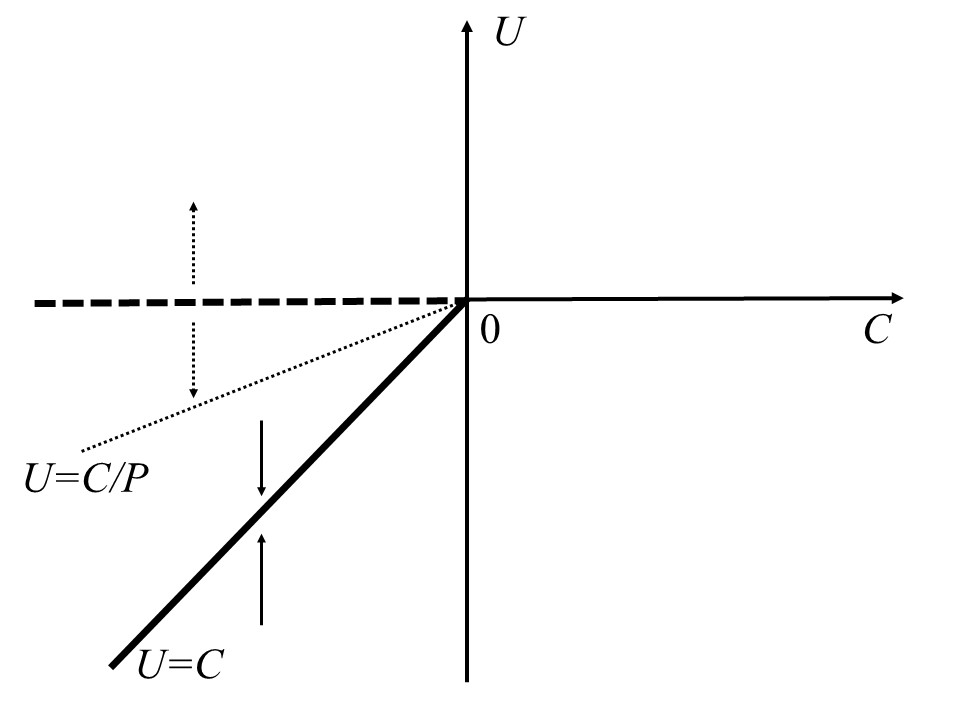}

\caption{\label{Fig.1D} The bifurcation diagram 
for the ultradiscrete saddle-node bifurcation generated from eq.(\ref{eqn:3-1}). }
\end{center}
\end{figure}

First let us consider the following  max-plus equation with the bifurcation parameter $C$:
	\begin{eqnarray}
		U_{n+1}=\max(PU_n, C).
	\label{eqn:3-1}
	\end{eqnarray}
Here, we set $P>1$.
Figure \ref{Fig.1} shows the graphs of eq.(\ref{eqn:3-1}) 
for (a) $C>0$, (b) $C=0$, and (c) $C<0$. 
For (a) $C>0$, the graph of eq.(\ref{eqn:3-1}) does not touch the diagonal $U_{n+1} = U_{n}$.
Then, eq.(\ref{eqn:3-1}) has no fixed point.
If $U_n<C/P$, $U_{n+1}=C$ and $U_{n+2}$ increases along $U_{n+2}=P U_{n+1} > U_{n+1}$.
For (b) $C=0$, eq.(\ref{eqn:3-1}) touches the diagonal at the origin of the graph.
Then, $U_n = 0$ is the only fixed point and it is half-stable.
For (c) $C<0$, the graph intersects the diagonal at the two points $U_n = 0$ and $U_n = C$, 
which are the unstable and stable fixed points, respectively. 
In fact, when $U_n \leq C/P$, $U_{n+1}=C$. 
When $C/P<U_n<0$, $U_n$ tends to $C/P$ first 
along $U_{n+1}=P U_{n}$, and after that $U_{n}$ finally arrives at $C$.
If $U_n>0$, $U_{n}$ goes to positive infinity along $U_{n+1}=P U_{n}$.
Then this bifurcation is saddle-node.
The bifurcation diagram is shown in Fig.\ref{Fig.1D}.
In the diagram, the solid arrows show the transition of $U_n$ 
to the stable point just at the next step.
The dotted arrows represent the transition satisfying $U_{n+1}=PU_n$.
%
We note that eq.(\ref{eqn:2-1-5}) is the special case 
of eq.(\ref{eqn:3-1}) with $P = 2$.
Then, Fig.\ref{Fig.1} (a), (b), and (c) correspond 
to graphical descriptions of {\bf 2.1(i), (ii)}, and {\bf (iii)} 
in Sec. \ref{sec:2-1}.
%
%
%
%
%
%

%
%
%
%
%
%
%
%

\subsection{Transcritical bifurcation}
\label{sec:3-2}

\begin{figure}[h!]
	\begin{center}
	\includegraphics[bb=0 0 960 720, width=5cm]{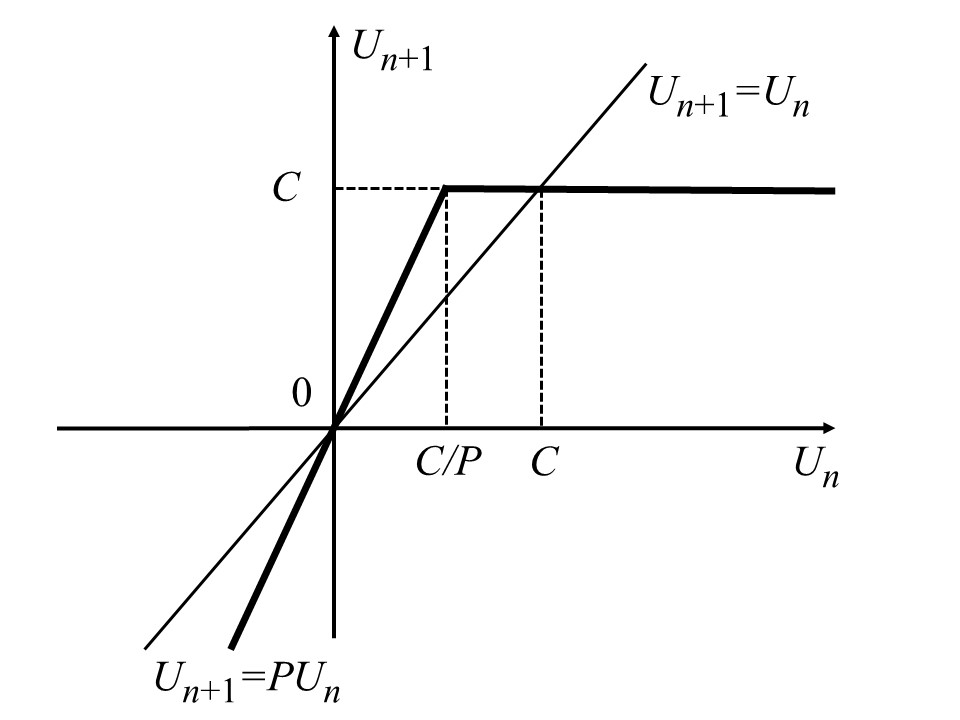}
	\hspace{0mm}
	\includegraphics[bb=0 0 960 720, width=5cm]{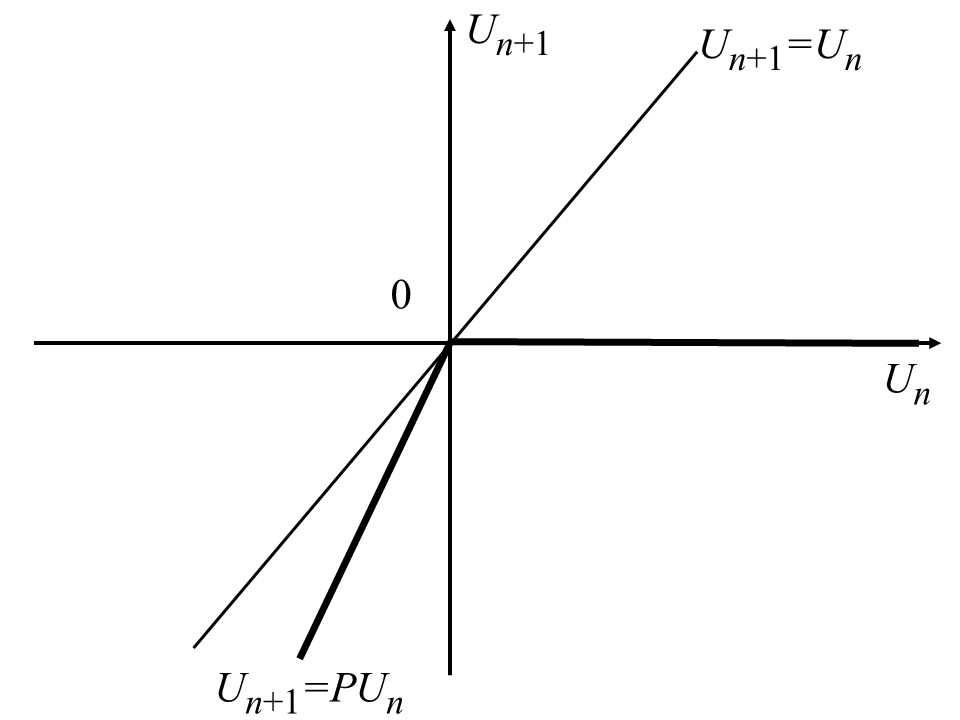}
	\hspace{0mm}
	\includegraphics[bb=0 0 960 720, width=5cm]{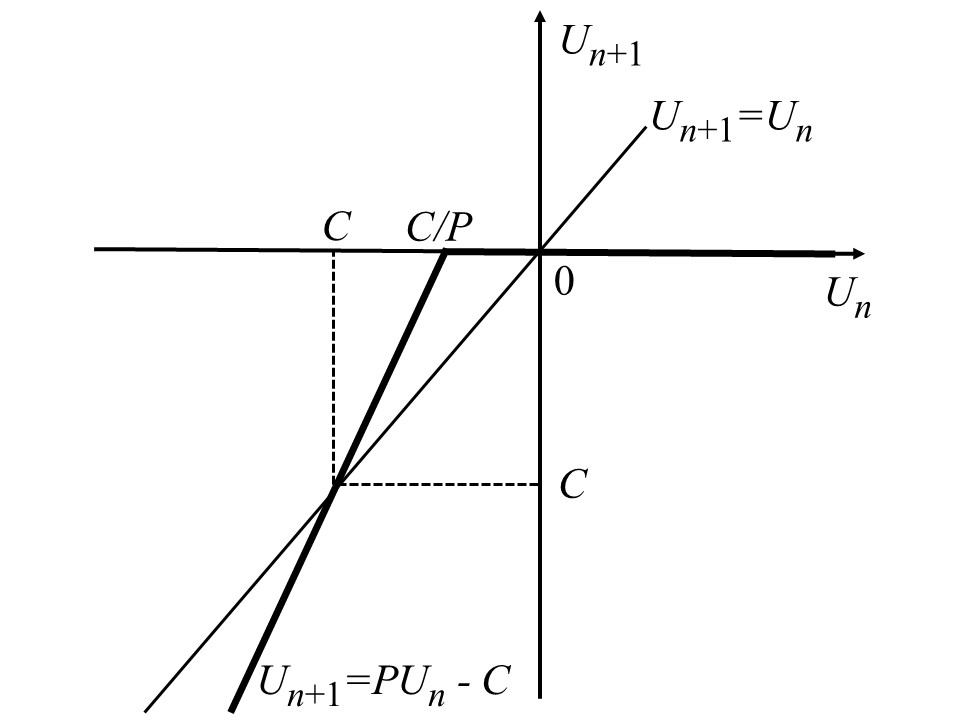}
	\\
	(a)
	\hspace{4cm}
	(b)
	\hspace{4cm}
	(c)
	\caption{\label{Fig.2} The graphs of eq.(\ref{eqn:3-2}) 
	where (a) $C>0$, (b) $C=0$, and (c) $C<0$.}
	\end{center}
\end{figure}

\begin{figure}[h!]
\begin{center}
\includegraphics[bb=0 0 960 720, width=5.5cm]{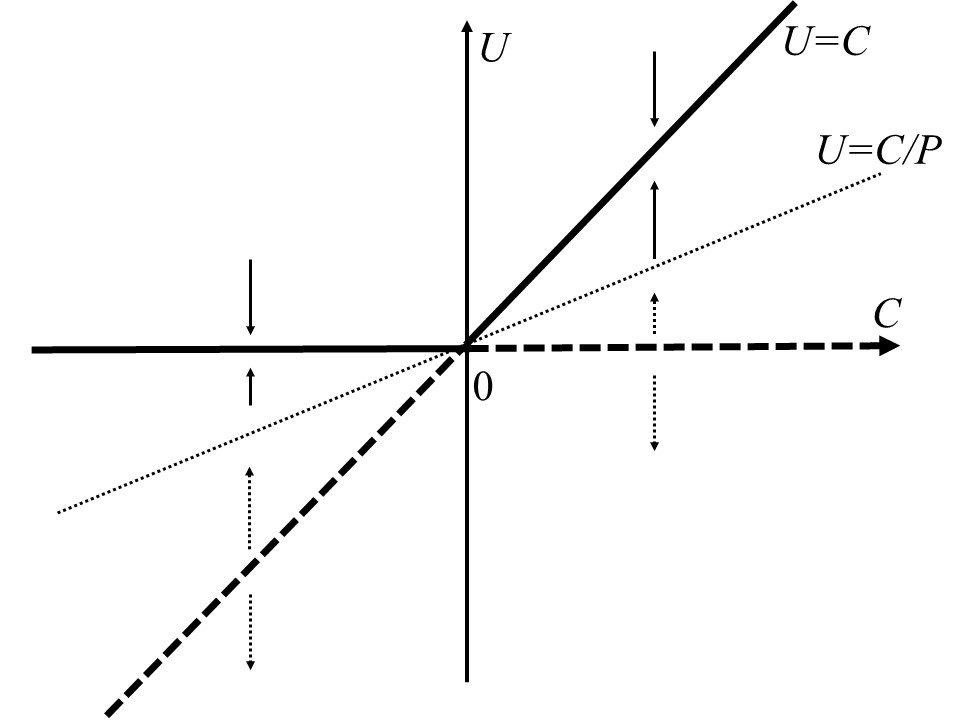}
\caption{\label{Fig.2D} The bifurcation diagram of eq.(\ref{eqn:3-2}).}
\end{center}
\end{figure}

Next we focus on the following max-plus equation 
with the bifurcation parameter $C$: 
	\begin{eqnarray}
		U_{n+1}=PU_n+\max(0,C)-\max(PU_n, C),
	\label{eqn:3-2}
	\end{eqnarray}
where $P>1$. 
The graphs of eq.(\ref{eqn:3-2}) with three different cases of $C$ 
are shown in Fig.\ref{Fig.2}.
For (a) $C>0$, the graph intersects the diagonal 
at the two points $U_{n}=0$ and $U_{n}=C$, 
which are unstable and stable fixed points, respectively.
Actually if $U_n<0$, $U_{n+1}=PU_{n}<U_n$ goes to negative infinity. 
If $0 < U_n \leq C/P$, there is the iteration step $m$, at which $U_m \geq C/P$ and $U_{m+1}=C$. 
If $U_n > C/P$, $U_{n+1} = C$.
For (b) $C=0$, there exists only one half-stable fixed point $U_{n}=0$; 
if $U_n>0$, $U_{n+1}=0$ and if $U_{n} < 0$, $U_{n+1}=PU_n (<U_n)$.
For (c) $C<0$, the graph of eq.(\ref{eqn:3-2}) intersects the diagonal 
at the two points $U_{n}=0$ (stable) and $U_{n}=C$ (unstable).
Figure \ref{Fig.2D} shows the bifurcation diagram; 
the bifurcation occurs at $C=0$.
%
%

In the case of ultradiscrete transcritical bifurcation 
shown by eq.(\ref{eqn:2-2-5}), 
the same piecewise linear graph as the case 
of eq.(\ref{eqn:3-2}) can be drawn by putting $P=2$.
In fact, Fig.\ref{Fig.2} (a), (b), and (c) become 
graphical representations 
of {\bf 2.2(i)}, {\bf (ii)}, and {\bf (iii)} 
in Sec. \ref{sec:2-2}, respectively.
%

%
%

\subsection{Supercritical pitchfork bifurcation}
\label{sec:3-3}

\begin{figure}[h!]
\begin{center}
\includegraphics[bb=0 0 960 720, width=5cm]{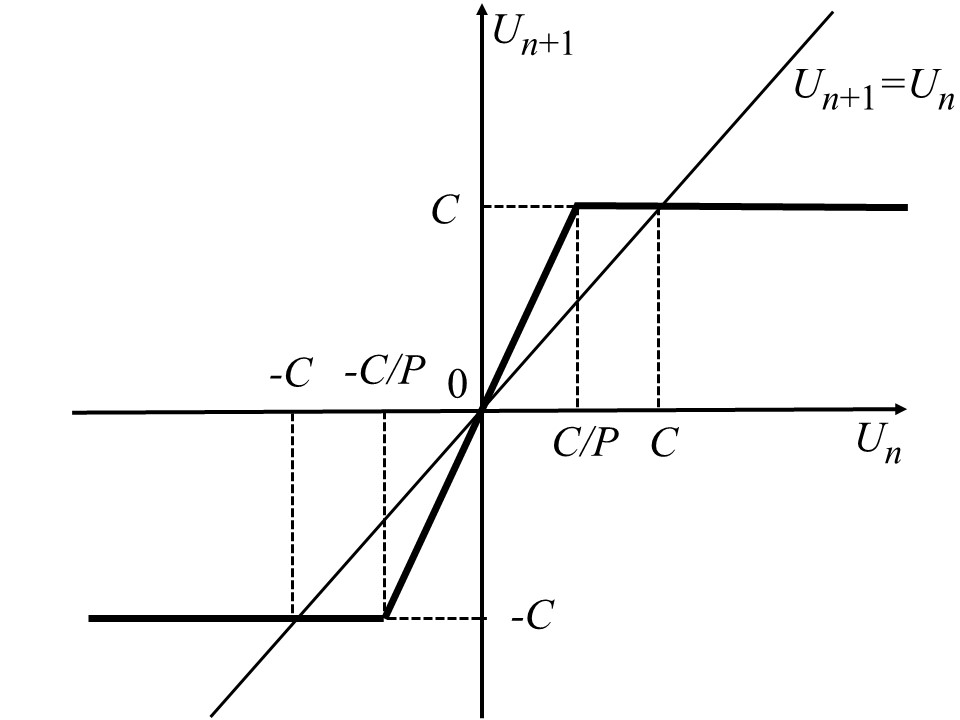}
\hspace{0mm}
\includegraphics[bb=0 0 960 720, width=5cm]{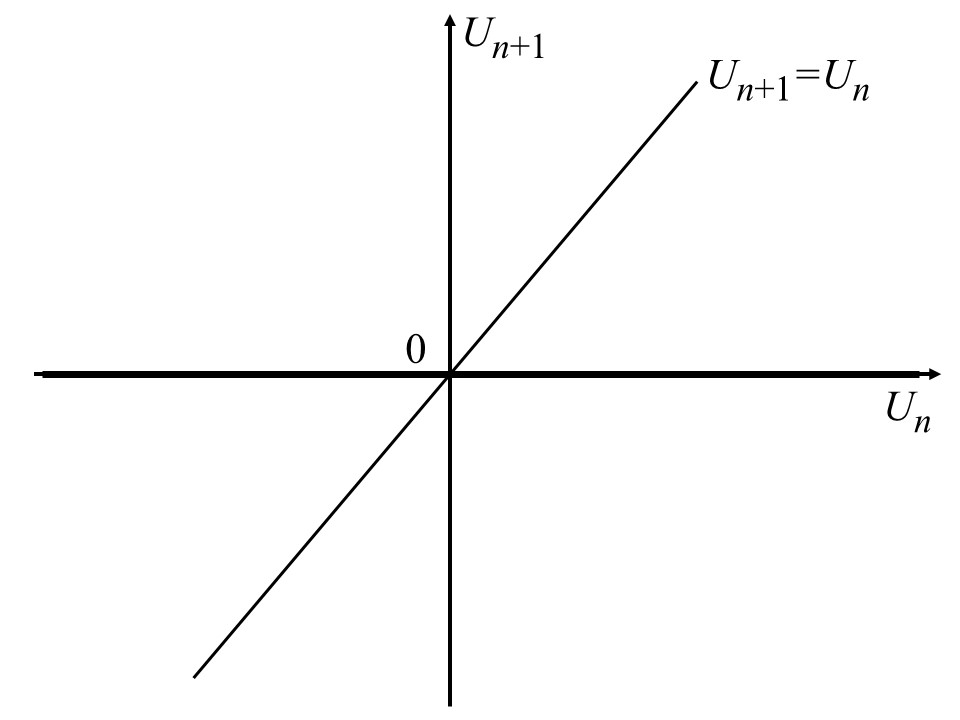}
\hspace{0mm}
\includegraphics[bb=0 0 960 720, width=5cm]{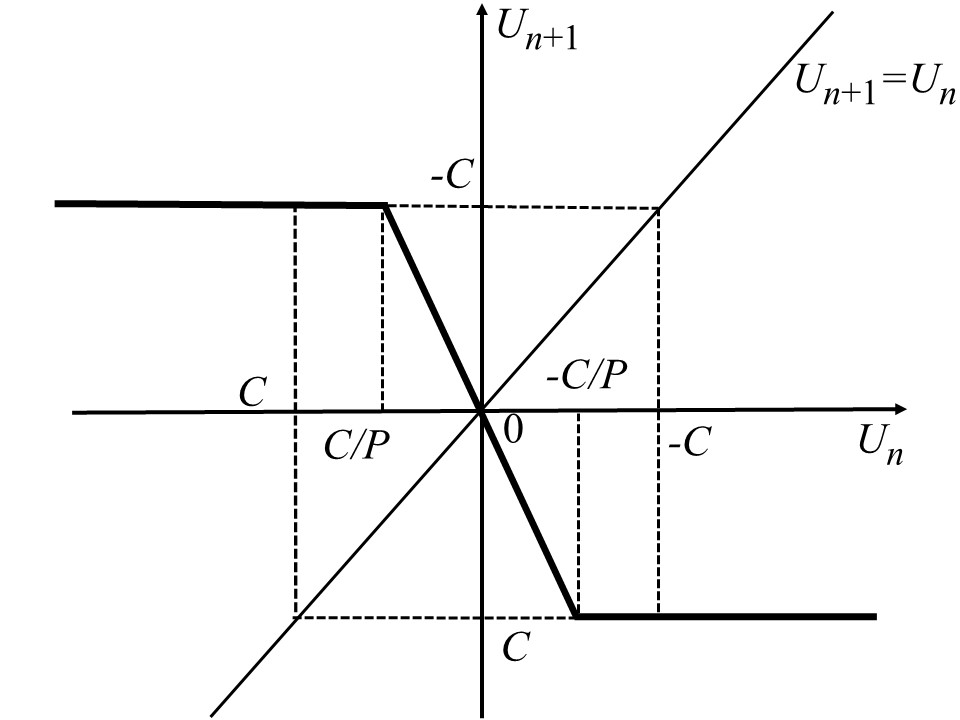}
\\
(a)
\hspace{5cm}
(b)
\hspace{5cm}
(c)

\caption{\label{Fig.3} The graphs of eq.(\ref{eqn:3-3}). (a) $C>0$, (b) $C = 0$, and $C<0$.}
\end{center}
\end{figure}

Here we consider the following max-plus equation: 
	\begin{eqnarray}
		U_{n+1}=\max\{PU_n+C,0\}-\max\{PU_n, C\},
	\label{eqn:3-3}
	\end{eqnarray}
where $P>1$. 
If we set $P=2$, eq.(\ref{eqn:3-3}) is the same 
as eq.(\ref{eqn:2-3-5}).
Figure \ref{Fig.3} (a), (b), and (c) show the graphs of eq.(\ref{eqn:3-3}) 
for $C > 0$, $C=0$, and $C < 0$, respectively.
Note that $U_{n+1}$ of eq.(\ref{eqn:3-3}) is an odd function of $U_{n}$
as shown in Fig.\ref{Fig.3}.
For (a) $C>0$, the graph of eq.(\ref{eqn:3-3}) intersects the diagonal 
at the three fixed points $U_{n} = 0, \pm C$.
If $U_n \geq C/P$ $(U_n \leq - C/P)$, $U_{n+1} = +C$ $(U_{n+1} = -C)$.
If $-C/P < U_n < 0$, there exists a certain $m$ at which $U_m \leq -C/P$ and $U_{m+1}=-C$.
In the same way,  $0 < U_n < C/P$ finally goes to $C$.
Therefore, $U_n = \pm C$ are stable and $U_n = 0$ is unstable.
As $C$ decreases, the two stable fixed points 
$U_n = \pm C$ merge when $C=0$ as shown in Fig.\ref{Fig.3}(b).
Then, $U_n = 0$ is only the stable fixed point for $C = 0$;
$U_1 = 0$ for any initial $U_0$.
We note that the dynamical properties 
of the cases (a) and (b) in Fig.\ref{Fig.3} 
correspond to {\bf 2.3(i)} and {\bf (ii)}, respectively.

\subsection{Flip bifurcation}
\label{sec:3-4}

For Fig.\ref{Fig.3}(c) where $C<0$ in eq.(\ref{eqn:3-3}), 
$U_n = 0$ becomes the unstable fixed point 
and there is a cycle $\mathcal{C} = \{+C,-C\}$ with period 2, 
which is the attractor of $U_n$.
The properties of $\mathcal{C}$ are shown as follows.
(i) $\mathcal{C}$ surrounds the unstable fixed point: $U_{n} = 0$.
(ii) Whenever $|U_0|>C$, $U_{n} \in \mathcal{C}$ for any $n\geq 1$.
(iii) If $U_0$ satisfies $0<|U_0|<C$, $U_n$ leaves 
from $0$ oscillating around $0$ 
and reaches the point $U_m$ such that $C/P \leq |U_m|$.
After that, $U_{n} \in \mathcal{C}$ for $n > m$.
Therefore, any $U_{n}$ starting from $U_0 \not =0$ is finally absorbed by the cycle $\mathcal{C}$.
%
%
%
%
The dynamical property shown in Fig.\ref{Fig.3}(c) 
corresponds to {\bf 2.3(iii)}. 
%
%
%
%
%



\section{Discussion}
\label{sec:4}


Regarding the derivation of eq.(\ref{eqn:2-3-5}) from eq.(\ref{eqn:2-3-1}), 
the dynamical property {\bf 2.3(iii)} 
of eq.(\ref{eqn:2-3-5}) for $C<0$ is not consistent 
with eq.(\ref{eqn:2-3-1}) for $c<1$.  
The dynamical inconsistency for $c<1$ can be resolved 
as follows.
For $c<1$, setting $\eta = 1-c~(>0)$, eq.(\ref{eqn:2-3-1}) becomes
	\begin{eqnarray}
	\frac{du}{dt} & = & 3(1-\eta)u(u-1)-u^3+1 \nonumber \\
						 & = & 	-(u-1)^{3} -3\eta u(u-1).
	\label{eqn:4-a-1}
	\end{eqnarray}
Then, we obtain the following discrete equation 
of eq.(\ref{eqn:4-a-1}) by the tropical discretization.
	\begin{eqnarray}
		u_{n+1} & = & \displaystyle
                  \frac{u_n+{\Delta t}\{ 3 u_n^2 + 3\eta u_n+1\}}
                  {1+{\Delta t}\{(u_n)^2 + 3\eta u_n+3\}}.
                  	\label{eqn:4-a-2}
	\end{eqnarray}
Applying the transformations 
\begin{equation}
	\Delta t = e^{T/\varepsilon},~~~ u_n = e^{U_n/\varepsilon},
	~~~ \eta =e^{E/\varepsilon }, 
\label{eqn:4-a-3}
\end{equation}
and assuming $T\geq 0$, 
the following ultradiscrete equations for $C<0$ can be obtained from eq.(\ref{eqn:4-a-2}).
	\begin{eqnarray}
		U_{n+1}  =  \max\{2U_n , E+U_n,0\}-\max\{2U_n , E+U_n,0\}  =  0.
		\label{eqn:4-a-3}
	\end{eqnarray}
Thus, for retaining the dynamical properties 
of the original equation for supercritical bifurcation,
we should consider eq.(\ref{eqn:2-3-5}) for $C \geq 0$ 
and eq.(\ref{eqn:4-a-3}) for $C < 0$.
%
%
%
It is important to properly select a form 
for the discretized equation of the original normal forms 
for retaining their original dynamical properties.

\begin{figure}[h!]
	\begin{center}
		\includegraphics[bb=0 0 960 720, width=5.5cm]{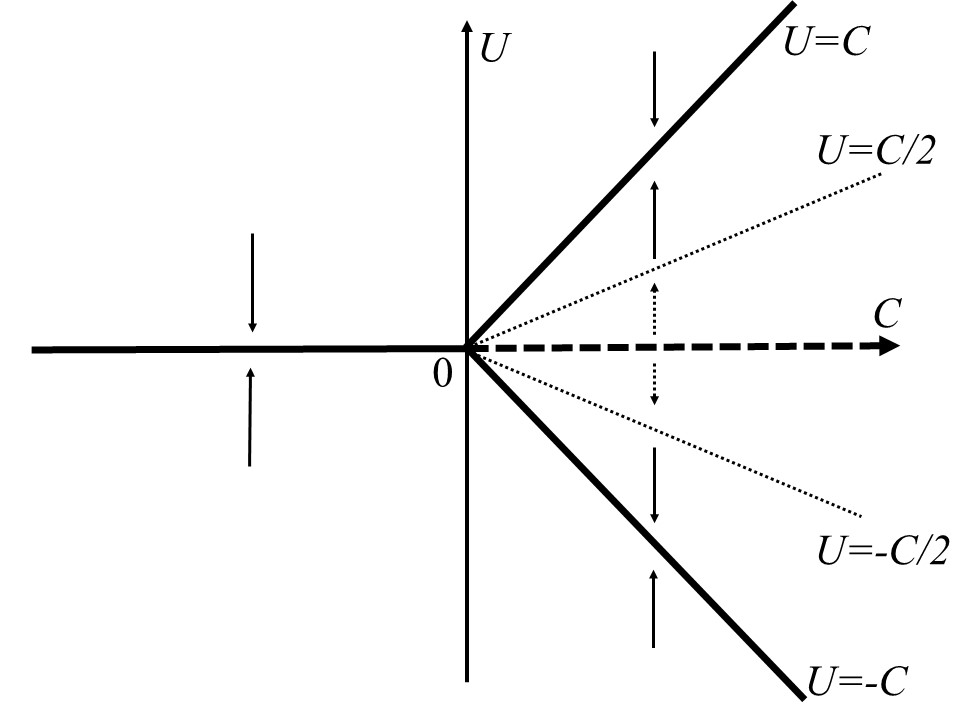}
		\caption{\label{Fig.3D} The bifurcation diagram 
		of eq.(\ref{eqn:4-1-5-1}). 
		The solid arrows show the transition of $U_n$ 
		to the stable point just at the next step.
		The dotted arrows represent the transition 
		satisfying $U_{n+1}=2U_n$.}
	\end{center}
\end{figure}

For ultradiscretization of supercritical bifurcation, 
now we start from eq.(\ref{eqn:2-1c}) as an another case 
instead of eq.(\ref{eqn:2-3-0}).
%
%
%
Applying the variable transformation $u \to u-1$ 
and setting $r = c-1$ to eq.(\ref{eqn:2-1c}), we obtain 
\begin{eqnarray}
	\frac{du}{dt} & = & (c - 1)(u-1)-(u-1)^3 \nonumber \\
						 & = & \begin{cases}
										-(u-1)(u-\alpha)(u-\gamma) \;\;\;\; \mbox{for  } c \geq 1, \\
										-(u-1)^{3} -\eta(u-1) \;\;\;\; \mbox{for  } c < 1,  \end{cases}
	\label{eqn:4-1-1}
\end{eqnarray}
where $\alpha=1 - \sqrt{c-1}$, $\gamma =1 + \sqrt{c-1}$, and $\eta = 1-c$.
Now we assume $\alpha > 0$.
%
%
Then, the discrete equation of eq.(\ref{eqn:4-1-1}) 
by the tropical discretization becomes 
	\begin{eqnarray}
		u_{n+1} & = & \displaystyle
                  \frac{u_n+{\Delta t}\{(1+\alpha + \gamma)u_n^2+\alpha \gamma \}}
                  {1+{\Delta t}\{(u_n)^2 + \alpha + \gamma+\alpha \gamma \}}
                  \;\;\;\; \mbox{for  } c \geq 1, 
		\label{eqn:4-1-2-1} \\
		u_{n+1} & = & \displaystyle
                  \frac{u_n+{\Delta t}\{ 3 u_n^2 + 1 + \eta \}}
                  {1+{\Delta t}\{(u_n)^2 + 3 + \eta\}}
                  \;\;\;\; \mbox{for  } c < 1.
		\label{eqn:4-1-2-2}
	\end{eqnarray}
Setting 
	\begin{equation}
		\begin{cases}
			\Delta t = e^{T/\varepsilon},~~~ u_n = e^{U_n/\varepsilon},~~~ c =e^{C/\varepsilon},  \\
			\alpha  = e^{A/\varepsilon},~~~ \gamma =e^{\Gamma/\varepsilon}, ~~~ \eta =e^{E/\varepsilon }, 
		\end{cases}
	\label{eqn:4-1-3}
	\end{equation}
%
%
and assuming $T\geq \max\{0,-(A+\Gamma)\}$, 
the following ultradiscrete equations are obtained from 
eq.(\ref{eqn:4-1-2-1}) and eq.(\ref{eqn:4-1-2-2}).
	\begin{eqnarray}
		U_{n+1} & = & \max\{2U_n+\Gamma, A+\Gamma \}-\max\{2U_n,\Gamma \}, 
                  \;\;\;\; \mbox{for  } C \geq 0, 
		\label{eqn:4-1-4-1} \\
		U_{n+1} 
					 & = & 0, 
                  \;\;\;\; \mbox{for  } C < 0.
		\label{eqn:4-1-4-2}
	\end{eqnarray}
(For derivation of eq.(\ref{eqn:4-1-4-1}) and eq.(\ref{eqn:4-1-4-2}), see Appendix \ref{sec:8-1}.)
Note that $A$ and $\Gamma$ depend on $C$ 
due to $c$ dependence of $\alpha$ and $\gamma$.
Furthermore when $C=0$, it is necessary 
that eq.(\ref{eqn:4-1-4-1}) is equal to eq.(\ref{eqn:4-1-4-2}), 
namely $A = \Gamma = 0$.
Then if we set 
%
		$A=-C$ and $\Gamma =C$,
%
eq.(\ref{eqn:4-1-4-1}) coincides with eq.(\ref{eqn:2-3-5}). 
%
%
In the present case, 
eq.(\ref{eqn:4-1-4-1}) is considered when $C \geq 0$, 
then we do not have to take the flip bifurcation 
shown in {\bf 2.3(iii)} into account.
If eq.(\ref{eqn:4-1-4-1}) and eq.(\ref{eqn:4-1-4-2}) 
are put in the following single equation, 
	\begin{equation}
				U_{n+1}  =  \max\{2U_n+\max(C, 0), 0\}-\max\{2U_n, C , 0\},
		\label{eqn:4-1-5-1} 
	\end{equation}
$C$-dependence of dynamical properties of eq.(\ref{eqn:4-1-5-1}) 
is shown in Fig.\ref{Fig.3D} 
as a bifurcation diagram.
Note that eq.(\ref{eqn:4-1-5-1}) holds for both cases 
which start from eq.(\ref{eqn:2-1c}) and eq.(\ref{eqn:2-3-0}).
The dynamical property of eq.(\ref{eqn:4-1-5-1}) is consistent 
with that of the original normal forms 
for supercritical pitchfork bifurcation.
%
%
Furthermore for $C \geq 0$, eq.(\ref{eqn:4-1-4-1}) is the same form 
as the ultradiscrete Allen-Cahn equation\cite{Murata2013} without diffusion effect.

\begin{figure}[h!]
\begin{center}
\includegraphics[bb=0 0 960 720, width=5cm]{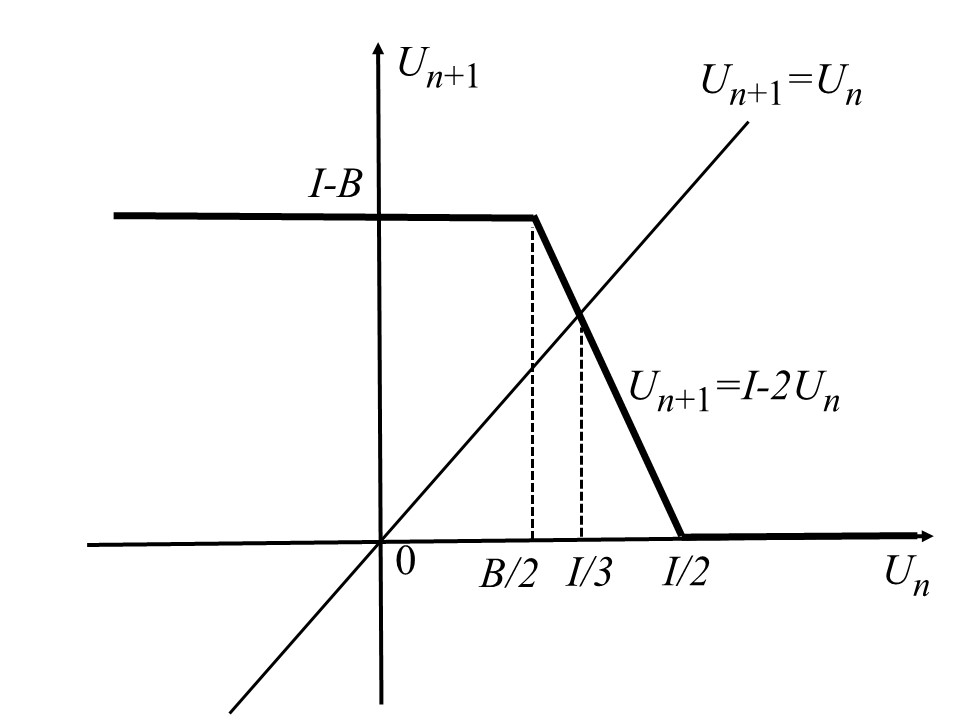}
\hspace{0mm}
\includegraphics[bb=0 0 960 720, width=5cm]{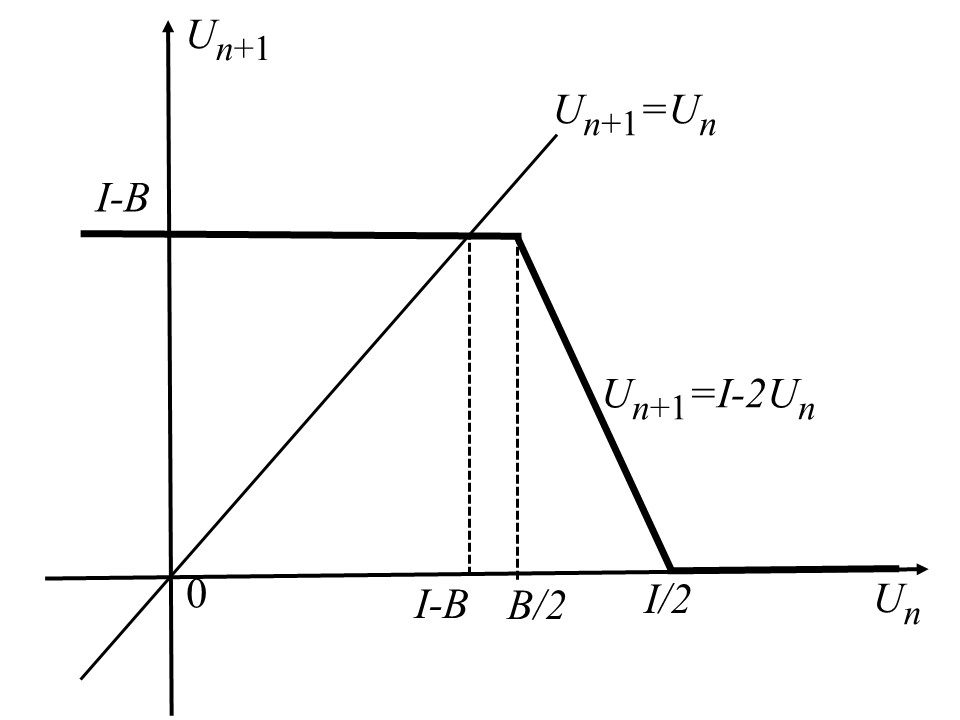}
\\
(a)
\hspace{4cm}
(b)\\
\vspace{3mm}
\includegraphics[bb=0 0 960 720, width=5cm]{EM3b.jpg}
\hspace{0mm}
\includegraphics[bb=0 0 960 720, width=5cm]{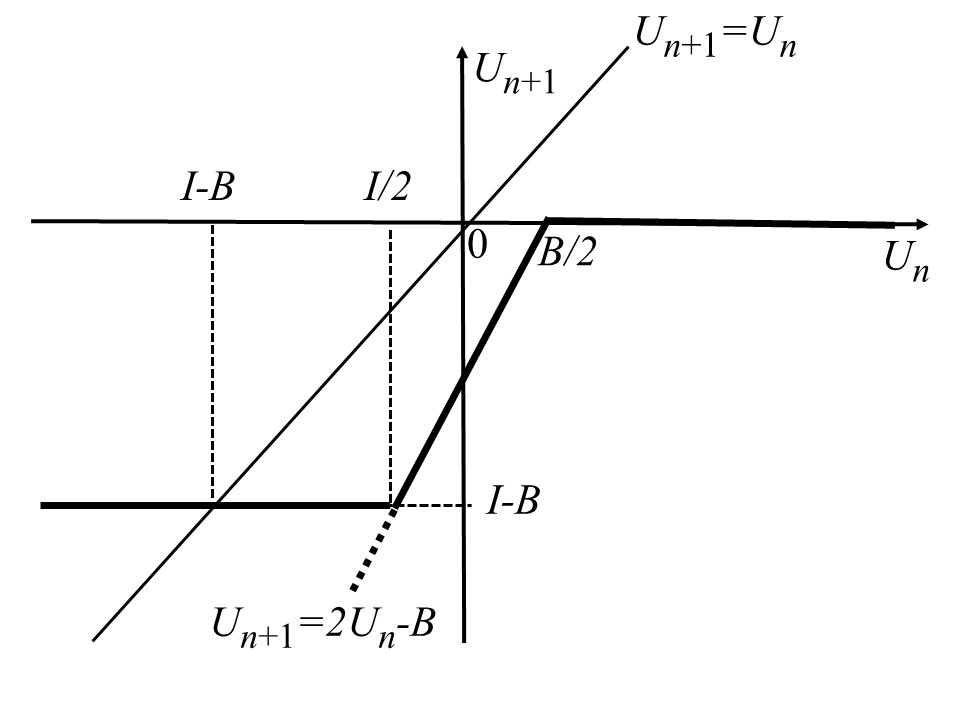}
\\
(c)
\hspace{4cm}
(d)\\
\caption{\label{Fig.5} The graphs of eq.(\ref{eqn:4-2-1}) 
with $B>0$. (a) $I>3B/2$, (b) $B<I\leq 3B/2$, (c) $I=B$, and (d) $I<B$.}
\end{center}
\end{figure}

Here, we focus on the ultradiscrete equation of FitzHugh-Nagumo model, 
which is given as the following reaction-diffusion system\cite{FNeq};
	\begin{eqnarray}{}
  		\displaystyle\frac{d u}{d t}
    	& = & u(1-u)(u-a)-v+i, 
		\nonumber \\
  		\displaystyle\frac{d v}{d t}
   		& = & \kappa u-\lambda v 
		\label{eqn:4-2-0}
	\end{eqnarray}
where $a, i, \kappa$, and $\lambda$ are positive 
and we set $0<a<1$. 
The ultradiscrete equation of FitzHugh-Nagumo model 
has been proposed by Sasaki et.al\cite{Sasaki2018}
as the following one-dimensional ultradiscrete equation 
	\begin{eqnarray}
		U_{n+1}=\max\{2U_n, I \}-\max\{2U_n, B\}.
	\label{eqn:4-2-1}
	\end{eqnarray}
By their numerical calculation, 
it was found that eq.(\ref{eqn:4-2-1}) has some characteristic solutions 
by changing the parameters $I$ and $B~(B>0)$.
%
%
For instance, a cyclic solution was found for $I>3B/2$.
In the graphical analysis, the dynamics of eq.(\ref{eqn:4-2-1}) for $B>0$ 
can be visualized as shown in Fig. \ref{Fig.5} (a)-(d).
Figure \ref{Fig.5} (a) shows the case of $I>3B/2$. 
We can find the cyclic solution $\mathcal{C}'$ surrounding the unstable fixed point $U_n = I/3$ 
which satisfies the properties of {\bf 3.4} stated above.
Then, $\mathcal{C}'$ is composed of the two points; 
$\mathcal{C}' = \{0,I-B\}$ for $I \geq 2B$,  
$\mathcal{C}' = \{2B-I,I-B\}$ for $I < 2B$.
Note that it is easily confirmed that eq.(\ref{eqn:4-2-1}) with $I=3B(>3B/2)$ 
is the same as eq.(\ref{eqn:3-3}) with $C=-B$ 
by the variable transformation $U_{n} \rightarrow U_{n}+B$. 
Note that if $U_0\geq I/2$, it is verified that $U_n=0$ when $n$ is odd and $U_n=I-B$ when $n$ is even.

\begin{figure}[h!]
\begin{center}
\includegraphics[bb=0 0 960 720, width=5cm]{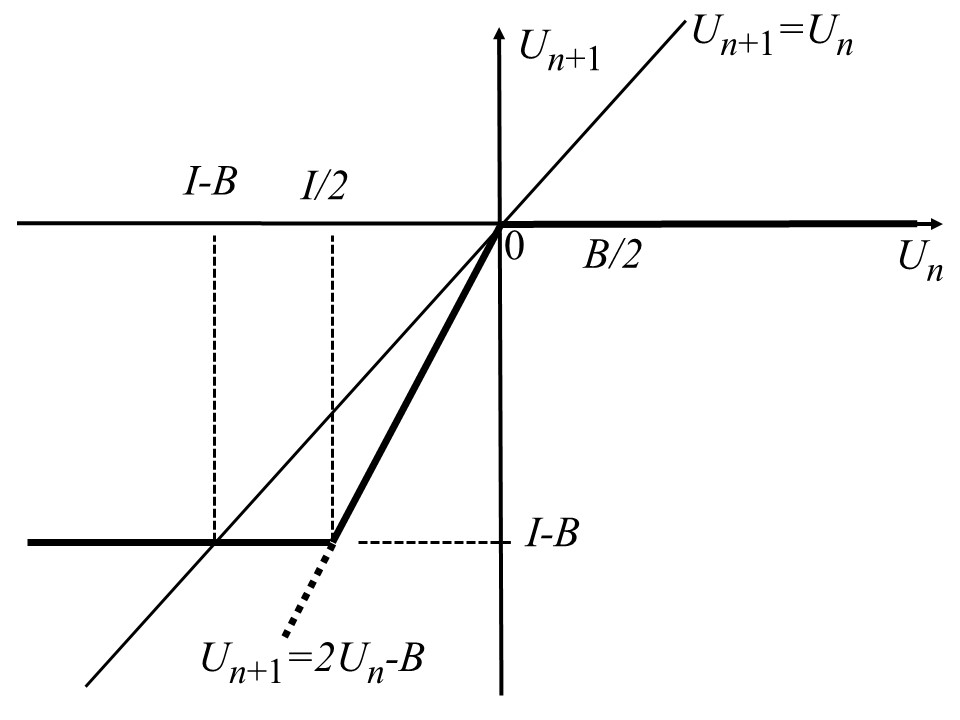}
\hspace{0mm}
\includegraphics[bb=0 0 960 720, width=5cm]{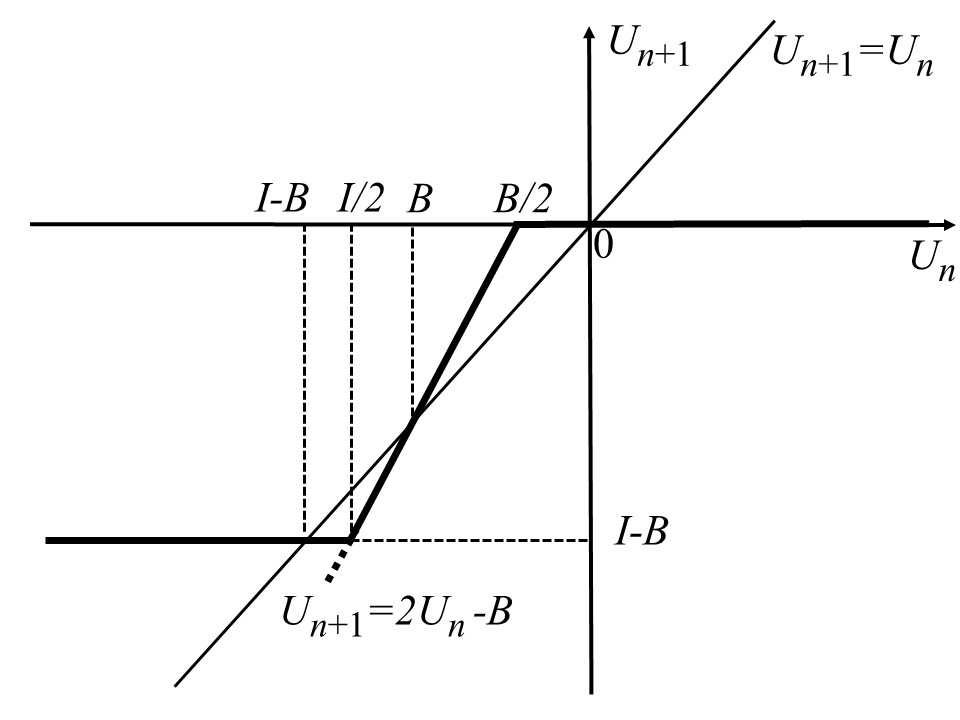}
\\
(a)
\hspace{4cm}
(b)\\
\vspace{3mm}
\includegraphics[bb=0 0 960 720, width=5cm]{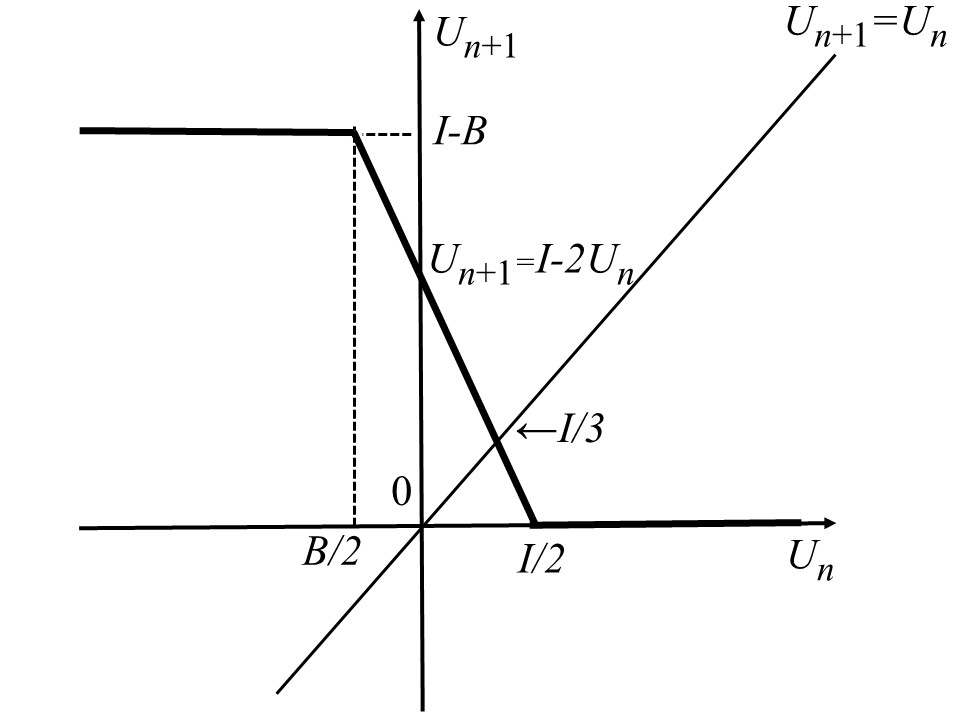}
\hspace{0mm}
\includegraphics[bb=0 0 960 720, width=5cm]{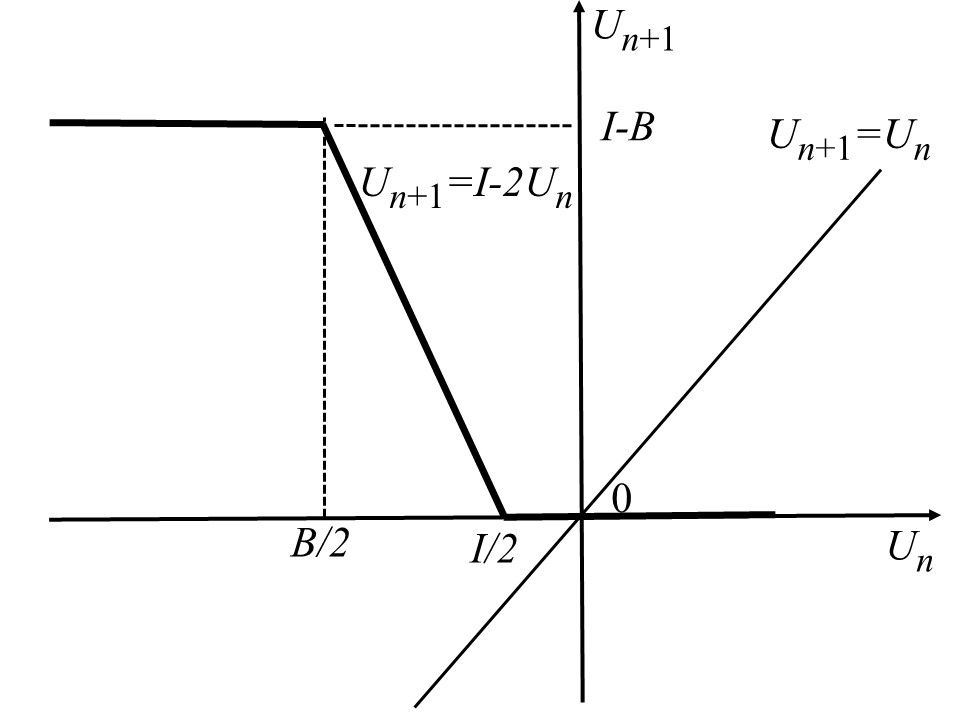}
\\
(c)
\hspace{4cm}
(d)\\

\caption{\label{Fig.6} The graphs of eq.(\ref{eqn:4-2-1}) with $B \leq 0$. (a) $I<B=0$, (b) $I<B<0$, (c) $B<0<I$, and (d) $B<I\leq 0$.}
\end{center}
\end{figure}

For $B<I\leq 3B/2$, 
Fig. \ref{Fig.5} (b) shows that $U_n = I-B$ is the only fixed point and it is stable.
According to the graphical analysis, when $U_0\leq B/2$, $U_1=I-B$.
On the other hand, when $U_0 > B/2$, $U_1 < I-B$,  and $U_2 = I-B$.
Note that this time evolution is similar to that of excitability.
In Figs. \ref{Fig.5} (c) and (d), all of the initial values finally go to the unique stable fixed point.
For $I=B$, Fig. \ref{Fig.5} (c) is the same as Fig. \ref{Fig.3} (b).
For $I<B$, it is confirmed that any $U_0$ finally arrives at the stable point $U_n = I-B$
according to eq.(\ref{eqn:4-2-1}); 
Fig. \ref{Fig.5} (d) shows the case of $I<0$. 
The graphical analysis can also be applied to eq.(\ref{eqn:4-2-1}) 
for $B<0$ as shown in Fig. \ref{Fig.6} (a)-(d).
In Fig.\ref{Fig.6} (a), $U_{n}=I-B$ and $U_{n}=0$ are stable and half-stable, respectively.
In Fig.\ref{Fig.6} (b), $U_{n}=I-B,U_{n}=0$ are stable and $U_{n}=B$ is unstable.
In Fig.\ref{Fig.6} (c), $U_{n}=I/3$ is the unstable fixed point, and a cyclic solution is obtained.
In Fig.\ref{Fig.6} (d), $U_{n}=0$ is stable in which the system produces 
the excitability like behavior when $I/2 \lesssim 0$.

Here, we comment the above cyclic solution of the ultradiscrete equation (\ref{eqn:4-2-1}).
It is noted that eq. (\ref{eqn:4-2-1}) is composed of one-variable discrete dynamics for $U_n$, 
although the original FitzHugh-Nagumo model (\ref{eqn:4-2-0}) shows the two-variable dynamics 
for $u$ and $v$.
This reduction of the variables occurs in the derivation of eq. (\ref{eqn:4-2-1}) 
from eq. (\ref{eqn:4-2-0}) through tropical discretization 
in which the variable $v$ in the first differential equation of eq.(\ref{eqn:4-2-0}) 
is discretized by $v_{n+1}$ instead of $v_{n}$ \cite{Sasaki2018}.
Since a one-dimensional differential equation can not have periodic solutions 
and the limit cycle solution of (\ref{eqn:4-2-0}) occurs 
due to the two-dimensional dynamics, 
the above periodic solution $\mathcal{C}'$ 
of eq. (\ref{eqn:4-2-1}) for $I>3B/2$ 
is not associated with the limit cyclic solution of the original FitzHugh-Nagumo model, 
but just caused by the discretization of eq.(\ref{eqn:4-2-0}).
Actually eq.(\ref{eqn:4-2-1}) can be also derived 
from the one dimensional differential equation of $u$, 
$\frac{d u}{d t}=u(1-u)(u-a)-\frac{\kappa}{\lambda}u+i$, 
which is obtained from eq.(\ref{eqn:4-2-0}) by setting $\frac{dv}{dt} =0$.  
Note that this periodicity is similar to the well-known relation 
between logistic map and logistic differential equation; 
the former has the periodic solution whereas the latter has no periodic solution.

Several developments of the current study are expected as the next steps.  
(i) The cases in higher dimensions, such as stability and bifurcations 
of simultaneous ultradiscrete equations\cite{Gibo2015,Ohmori2016}. 
(ii) Other topics for the one-dimensional discrete dynamical systems such as chaos\cite{Melo}.
(iii) Application to ultradiscrete equations with spatial variables.
%



\section{Conclusion}
\label{sec:5}

Ultradiscrete equations are derived from the normal forms 
of saddle node, transcritical, and supercritical pitchfork bifurcations.
These derived equations exhibit ultradiscrete bifurcations, 
namely the similar bifurcation properties to the original normal forms. 
From supercritical bifurcation, 
we encounter another ultradiscrete bifurcation, 
similar to the flip bifurcation, 
where there is a stable cycle around a unstable fixed point.
With the aid of graphical analysis, 
these ultradiscrete bifurcations can be characterized 
with the much simpler max-plus equations.
As an example of application of graphical analysis, 
we can grasp essential dynamical features 
of the ultradiscrete equation for FitzHugh-Nagumo model, 
which was previously proposed in \cite{Sasaki2018}.
%

%
%
%
%
%
\appendix

\numberwithin{equation}{section}
\makeatletter

\section{Derivation of the ultradiscrete equations}
\label{sec:6}
\subsection{Saddle-node bifurcation: eq.(\ref{eqn:2-1-4}) $ \rightarrow $ eq.(\ref{eqn:2-1-5}) }
\label{sec:6-1}

When we set $T\geq \max\{0,-C/2\}$, 
\[
T+ \max(2U_n, C)-U_n=T+C/2+\max(U_n-C/2,-U_n+C/2) \geq 0
\]
since $T+C/2\geq 0$ and $\max(U_n-C/2,-U_n+C/2) = |U_n-C/2| \geq 0$.
Hence, $\max\{U_n, T+\max(2U_n, C)\}=T+\max(2U_n,C)$, 
and eq.(\ref{eqn:2-1-5}) is obtained from eq.(\ref{eqn:2-1-4}).

\subsection{Transcritical bifurcation : eq.(\ref{eqn:2-2-3}) $ \rightarrow $ eq.(\ref{eqn:2-2-5})}
\label{sec:6-2}

When we set $T \geq -C/2$,
\[
T+ U_n + \max(0, C)-U_n=T+C/2+\max(-C/2,C/2) \geq 0
\]
since $T+C/2\geq 0$ and $\max(-C/2,C/2) = |C/2| \geq 0$.
Moreover, 
\[
T+ \max(2U_n, C)-U_n=T+C/2+\max(U_n-C/2,-U_n+C/2) \geq 0
\]
since $T+C/2\geq 0$ and $\max(U_n-C/2,-U_n+C/2) = |U_n-C/2| \geq 0$.
Thus, eq.(\ref{eqn:2-2-5}) is derived from eq.(\ref{eqn:2-2-3}) as follows:
    \begin{eqnarray}
      U_{n+1} & = &U_n + \max\{U_n, T+ U_n + \max(0, C)\}-\max\{U_n, T+\max(2U_n, C)\}
                  \nonumber \\
            & = & U_n+T+U_n+\max(0,C)-\{T+\max(2U_n,C)\}
                  \nonumber \\
            & = & 2U_n+\max(0,C)-\max(2U_n,C)
  \label{eqn:appenA-2}
    \end{eqnarray}

\subsection{Supercritical pitchfork bifurcation : eq.(\ref{eqn:2-3-4}) $ \rightarrow $ eq.(\ref{eqn:2-3-5})}
\label{sec:6-3}

When we set $T\geq \max\{-C/2,-C\}$, 
\[
T+ \max(2U_n+C, 0)-U_n=T+C/2+\max(U_n+C/2,-U_n-C/2) \geq 0
\]
since $T+C/2 \geq 0 $ and $\max(U_n+C/2,-U_n-C/2) =|U_n+C/2| \geq 0$.
Moreover, 
$T+\max(2U_n,C) = T+C+\max(2U_n-C, 0) \geq 0 $ from $T+C \geq 0$.
Thus, eq.(\ref{eqn:2-3-4}) 
\[
	U_{n+1} = \max\{U_n, T+ \max(2U_n+C, 0)\}-\max \{0, T+\max(2U_n,C)\}
\]
becomes
	\begin{eqnarray}
		U_{n+1}= T+ \max(2U_n+C, 0) - \{T+\max(2U_n,C)\},
	\label{eqn:appeA-3}
	\end{eqnarray}
and, eq.(\ref{eqn:2-3-5}) is obtained from eq.(\ref{eqn:appeA-3}).

\color{black}

\section{Supplementaries for dynamical properties of the ultradiscrete bifurcations}
\label{sec:7}

For simplicity, we set $n = 0$ without loss of generality.

\subsection{Saddle node bifurcation (sec.2.1)}
\label{sec:7-1}

\begin{description}
  \item[2.1(i)]
		For $C>0$, there is no fixed point.
		Actually if $U_0 \leq 0, U_1 = \max(2U_0, C)\geq C>0 \geq U_0$.
		If $U_0 > 0, U_1=\max(2U_0,C) \geq 2U_0>U_0$.
	\item[2.1(ii)]
		For $C = 0$, eq.(\ref{eqn:2-1-5}) becomes $U_1=\max(2U_0,0)$. 
		Then, $U_0 = 0$ is a fixed point.
		If $U_0 < 0, U_1=\max(2U_0,0)=0$.
		If $U_0>0, U_1=\max(2U_0,0)=2U_0>U_0$.  
		Hence, $0$ is half-stable.
	\item[2.1(iii)]
		For $C<0$, $U_0 = C$ and $U_0 = 0$ are fixed points.
		If $U_0 \leq C/2$, $U_1=C$.
		If $C/2<U_0<0, U_1=2U_0<U_0$.
		If $0<U_0, U_1=2U_0>U_0$.
		Hence, $U_0 = C$ and $U_0 = 0$ are stable and unstable, respectively.
\end{description}

\subsection{Transcritical bifurcation (sec.2.2)}
\label{sec:7-2}

\begin{description}
  \item[2.2(i)]
	For $C>0$, eq.(\ref{eqn:2-2-5}) becomes $U_1=2U_0+C-\max(2U_0, C)$.
	Then, $U_0 = 0$ and $U_0 = C$ are fixed points.
	If $U_0 < 0, U_1 = 2U_0+C-C=2U_0< U_0$.
	If $0<U_0<C/2, U_1=2U_0>U_0$.
	If $C/2 \leq U_0 <C, U_1=2U_0+C-2U_0=C$.
	If $C<U_0, U_1=C$.
	Hence, $C$ and $0$ are stable and unstable, respectively.
	\item[2.2(ii)]
	For $C=0$, it follows from eq.(\ref{eqn:2-2-5}) that $U_1=2U_0-2\max(U_0,0)$. 
	Then, $U_0=0$ is the only one fixed point.
	If $U_0<0$, $U_1=2U_0<U_0$.
	If $U_0>0$, $U_1=0$.
	Hence, $0$ is half-stable.
	\item[2.2(iii)]
	For $C<0$, eq.(\ref{eqn:2-2-5}) becomes $U_1=2U_0-\max(2U_0, C)$.
	Then, $U_0 = 0$ and $U_0 = C$ are fixed points.
	If $U_0 < C$, $U_1 = 2U_0-C< U_0$.
	If $C<U_0<C/2$, $U_1=2U_0-C>U_0$.
	If $U_0 \geq C/2$, $U_1=0$.
	Hence, $U_0 = 0$ and $U_0 = C$ are stable and unstable, respectively.
\end{description}

\subsection{Supercritical pitchfork bifurcation (sec.2.3)}
\label{sec:7-3}

\begin{description}
	\item[2.3(i)]
		For $C > 0$, the stability of eq.(\ref{eqn:2-3-5}) have already been studied\cite{Ohmori2016}.
		If $U_0 \leq -C/2$, $U_1=-C$.
		If $C/2 \leq U_0$, $U_1=C$.
		If $-C/2<U_0<0$, $U_1=2U_0<U_0$ and there is a finite time $m (>0)$ 
		at which $U_m \leq -C/2$ and $U_{m+1} = 0$.
		Similarly if $0<U_0<C/2$, 
		there is a finite time $m (>0)$ at which $C/2 \leq U_m$ and $U_{m+1} = C$.
		In conclusion, $C,-C$ are stable fixed points, 
		$0$ is a unstable fixed point.
  \item[2.3(ii)]
	For $C = 0$, it follows from eq.(\ref{eqn:2-3-5}) that $U_1=0$ for any $U_0$. 
	$0$ is a stable fixed point.
	\item[2.3(iii)]
		Set $C < 0$.
		If $U_0 \leq C/2$, $U_1=\max(2U_0+C,0)-C=\max(2U_0,-C)=-C$.
		If $C/2 < U_0 < 0$, $U_1=\max(2U_0+C,0)-2U_0=\max(C,-2U_0)=-2U_0>-U_0$.
		If $0 < U_0 < -C/2$, $U_1=-\max(2U_0,C)=-2U_0<-U_0$.
		If $-C/2 \leq U_0$, $U_1=2U_0+C-2U_0=C$.
		$0$ is a unstable fixed point.

\end{description}

\section{Supercritical pitchfork bifurcation : 
eqs.(\ref{eqn:4-1-2-1}) and (\ref{eqn:4-1-2-2}) $ \rightarrow $ 
eqs.(\ref{eqn:4-1-4-1}) and (\ref{eqn:4-1-4-2})}
\label{sec:8-1}

From eqs. (\ref{eqn:4-1-2-1}) and (\ref{eqn:4-1-2-2}) with eq. (\ref{eqn:4-1-3}), 
the following ultradiscrete equations are obtained 
through the ultradiscretization shown by eq.(\ref{eqn:1-1}):
\begin{eqnarray}
	U_{n+1} & = & \max\{U_n, T+ \max(2U_n+\Gamma, A+\Gamma)\}-\max\{0, T+\max(2U_n, \Gamma)\}, 
		\label{eqn:appenA-8-1} \\
	U_{n+1} & = & \max\{U_n, T+ \max(2U_n, 0, E)\}-\max\{0, T+\max(2U_n, 0, E)\}. 
		\label{eqn:appenA-8-2}
\end{eqnarray}
When we set $T \geq \max\{0,-A/2\}-\Gamma$ 
for eq.(\ref{eqn:appenA-8-1}), 
\[
T+ \max(2U_n+\Gamma, A+\Gamma)-U_n=T+\Gamma+A/2+\max(U_n-A/2,-U_n+A/2) \geq 0
\]
since $T+\Gamma+A/2 \geq 0 $ and $\max(U_n-A/2,-U_n+A/2) \geq 0$.
Hence, $\max\{U_n, T+ \max(2U_n+\Gamma, A+\Gamma)\} = T+ \max(2U_n+\Gamma, A+\Gamma)$.
Moreover, 
$\max\{0, T+\max(2U_n, \Gamma)\}=T+\max(2U_n,\Gamma)$ 
since $T+\max(2U_n, \Gamma) \geq T+\Gamma \geq 0$. 
Therefore, eq.(\ref{eqn:4-1-4-1}) is derived 
from eq.(\ref{eqn:appenA-8-1}).

When we set $T \geq -\max(0, E/2)$
for eq.(\ref{eqn:appenA-8-2}), 
\[ T+ \max(2U_n, 0, E)-U_n 
 =  T+\max(0, E/2)+\max\{U_n-\max(0, E/2),-U_n+\max(0, E/2)\} \geq 0 \nonumber 
\]
since $T+\max(0, E/2) \geq 0$ 
and $\max\{U_n-\max(0, E/2),-U_n+\max(0, E/2)\} \geq 0$.
Hence, $\max\{U_n, T+ \max(2U_n, 0, E)\}=T+ \max(2U_n, 0, E)$. 
Moreover, 
$\max\{0, T+\max(2U_n, 0, E)\}=T+\max(2U_n, 0, E)$ 
from $T+ \max(2U_n, 0, E) \geq T+\max(0,E) 
\geq T+\max(0,E/2) \geq 0$.
Thus, eq.(\ref{eqn:4-1-4-2}) is derived from eq.(\ref{eqn:appenA-8-2}).

\bigskip

\noindent
{\bf Acknowledgement}

The authors are grateful to Prof. D. Takahashi, 
Prof. T. Yamamoto, and Prof. Emeritus A. Kitada 
at Waseda university for useful suggestions and encouragements. 
Also, we greatly appreciate the valuable comments 
from an reviewer.

\bigskip

\noindent
{\bf Data Availability}

Data sharing is not applicable to this article as no new data were created or analyzed in this study.\\

\end{document}